# The Parameters of Menzerath-Altmann Law in Genomes[*]

Jaume Baixeries[1], Antoni Hernández-Fernández[1,2], Núria Forns[3] & Ramon Ferrer-i-Cancho[1]

[1]Complexity and Quantitative Linguistics Lab. Departament de Llenguatges i Sistemes Informàtics. TALP Research Center/LARCA. Universitat Politècnica de Catalunya. Barcelona (Catalonia), Spain.

[2]Departament de Lingüística General. Universitat de Barcelona. Barcelona (Catalonia), Spain.

[3]Departament de Microbiologia, Facultat de Biologia, Universitat de Barcelona. Barcelona (Catalonia), Spain.

## ABSTRACT

The relationship between the size of the whole and the size of the parts in language and music is known to follow Menzerath-Altmann law at many levels of description (morphemes, words, sentences…). Qualitatively, the law states that larger the whole, the smaller its parts, e.g., the longer a word (in syllables) the shorter its syllables (in letters or phonemes). This patterning has also been found in genomes: the longer a genome (in chromosomes), the shorter its chromosomes (in base pairs). However, it has been argued recently that mean chromosome length is trivially a pure power function of chromosome number with an exponent of -1. The functional dependency between mean chromosome size and chromosome number in groups of organisms from three different kingdoms is studied. The fit of a pure power function yields exponents between -1.6 and 0.1. It is shown that an exponent of -1 is unlikely for fungi, gymnosperm plants, insects, reptiles, ray-finned fishes and amphibians. Even when the exponent is very close to -1, adding an exponential component is able to yield a better fit with regard to a pure power-law in plants, mammals, ray-finned fishes and amphibians. The parameters of Menzerath-Altmann law in genomes deviate significantly from a power law with a -1 exponent with the exception of birds and cartilaginous fishes.

# 1. INTRODUCTION

Menzerath–Altmann law is a linguistic law relating *y*, the size of a construct (e.g. a word), with *x* the size of its constituents (e.g., syllables). From a qualitative point of view, the law states that the larger the size of a construct the smaller the parts (Altmann, 1980). For instance, the longer a word (in syllables), the shorter its syllables (in letters or phonemes). The law is mathematically defined through the equation (Altmann, 1980)

$$y = ax^b e^{cx}, \qquad (1)$$

where *a*, *b* and *c* are parameters. The law bears the name of the researcher who observed the qualitative dependence between the size of the whole and the size of the parts in language (P. Menzerath) and the researcher who put it into mathematical form (G. Altmann). The mathematical function in Eq. 1 has been used not only in quantitative linguistics but also in other studies of scaling laws of genomes (Molina & van Nimwegen, 2009). Patterning consistent with Menzerath-Altmann law in the genomes of various groups of organisms has been reported (Ferrer-i-Cancho & Forns, 2009) by taking *y* as the size of a genome in chromosomes ($L_g$) and taking *x* as the length of its chromosomes in million base pairs ($L_c$). These analyses made two major simplifications:

- *x* was taken as a mean length for consistency with previous linguistic research (e.g., Boroda & Altmann, 1991) and references therein) and due to the absence of information about the length of concrete chromosomes in public database for a sufficiently large number of organisms.
- Agreement with Menzerath-Altmann law not through a fit of Eq. 1 but through a statistical correlation test. A negative correlation between *y* and *x* was considered as consistent with the law but it does not imply that Eq. 1 actually holds.

It has been argued that both simplifications could lead to trivial results (Solé, 2010). In (Ferrer-i-Cancho & Forns, 2009), chromosome lengths were defined as an average, i.e. $L_c = G/L_g$ for a given organism of genome length size *G* (in base pairs) and $L_g$ chromosomes. It has been argued that the definition $L_c = G/L_g$ for a certain organism implies $L_c \sim 1/L_g$ for any organism and thus the negative dependency between $L_c$ and $L_g$ reported by (Ferrer-i-Cancho & Forns, 2009) is unavoidable and therefore not relevant (Solé, 2010). By analyzing only two groups of the eleven major groups of organisms that were considered in the original research on genomes (Ferrer-i-Cancho & Forns, 2009), it was concluded that the negative correlations between $L_c$ and $L_g$ reported in the study by Ferrer-i-Cancho & Forns (2009) are "an inevitable consequence of the definitions of chromosome and genome" (Solé, 2010).

In this article, the dependency between mean chromosome length and chromosome number is studied with statistical depth. Three possible nested mathematical definitions for the dependency between $L_c$, the mean chromosome length and $L_g$, the genome size (in base pairs), are considered:

$$L_c = a L_g^{-1} \qquad (2)$$

$$L_c = aL_g^b \tag{3}$$

$$L_c = aL_g^b e^{cL_g}, \tag{4}$$

where *a*, *b* and c are constants. Eq. 2 is the one that, according to Solé (2010), genomes must obey trivially when chromosome length is defined as a mean. Eq. 4 is an adaptation of Eq. 1 to genomes. If the arguments by Solé (2010) were correct, the fit of Eq. 3 to genome data should give $b \approx -1$ and the fit of Eq. 4 should give $b \approx -1$ and $c \approx 0$. Incidentally, *b* = -0.6 is reported for the fit of Eq. 3 to ants (*N* = 105) in the pioneering work by Wilde & Schwibbe (1989). Here the analysis of Menzerath-Altmann law will be extended to the large groups of organisms employed in recent studies (Ferrer-i-Cancho & Forns, 2009; Hernández-Fernández et al, 2011) by means of the three nested models defined in Eqs 2-4.

## 2. RESULTS

Table 1 shows that the fit of Eq. 3 gives values of the parameter *b* that vary between -1.45 (jaw-less fishes) and 0.1 (amphibians). Table 1 also indicates that *b* = -1 is unlikely for fungi, gymnosperm plants, insects, reptiles, jawless fishes, ray-finned fishes and amphibians.

Table 2 indicates that Eq. 4 gives a significantly better fit than Eq. 3 for plants, mammals, ray-finned fishes and amphibians. Table 3 indicates that Eq. 2 gives always a poorer fit except for birds, mammals and cartilaginous fishes. However, Table 3, shows that the fit of Eq. 4 is always better than that of Eq. 3 in all groups except insects, birds, cartilaginous fishes and jawless fishes, where the fit is worse. The latter is interpreted as a failure of the fitting algorithm because Eq. 3 is a particular case of Eq. 4 with *c* = 0.

## 3. DISCUSSION

Table 4 summarizes all the qualitative results obtained so far on the dependency between $L_g$ and $L_c$ in this article and previous research in our major groups of organisms. The fact that the correlation between $L_g$ and $L_c$ is not significant but Eq. 4 gives a better fit in gymnosperm plants and ray-finned fishes suggests that their dependency between $L_g$ and $L_c$ may not be monotonic as the parameters reported for Eq. 4 in Table 2 indicate.

It has been argued that the definition of $L_c$ as a mean, namely $L_c = G/L_g$ implies an inverse proportionality dependency between $L_g$ and $L_c$, i.e. $L_c \sim 1/L_g$ (Solé, 2010). We have examined three nested models and shown that Eqs. 3 and 4 are able to approximate the actual dependency between mean chromosome length and chromosome number better than the inverse proportionality relationship (Eq. 2), with the only exception of birds and cartilaginous fishes (Table 4). None of these two groups coincides with the two groups selected by Solé (2010), i.e. mammals and plants to support his claim that Menzerath-Altmann law in genomes is a trivial power law with a -1 exponent. Notice that the results reported here do not imply that $L_c \sim 1/L_g$ gives a perfect fit for birds and cartilaginous fishes. Although more powerful mathematical and statistical arguments have been used to discard it (Baixeries et al, 2011; Hernández-Fernández et al, 2011), these two groups should be the subject of future research.

Our study of the dependency between mean chromosome length and chromosome number has been focused on a small family of nested mathematical models motivated by the proposal of an inverse proportionality relationship by Solé (2010), the mathematical definition of Menzerath-Altmann law in quantitative linguistics research (Altmann 1980, Teupenhayn & Altmann, 1984) and research on scaling laws in genomes (Molina & van Nimwegen, 2009). However, the issue of the mathematical function that would give a priori the best fit needs to be investigated further. Our analysis does not exclude the possibility that there are more appropriate functions to describe such dependency. With this regard, the motivation of the simple correlation analysis by Ferrer-i-Cancho & Forns (2009) was staying as much neutral as possible about the actual dependency between mean chromosome length and chromosome number. A trivial correlation can arise if genome size is statistically independent from chromosome number (Baixeries et al, 2011), a property that has been rejected for the majority of groups (Hernández-Fernández, 2011). The combination of a simple correlation analysis with further analyses to exclude trivial sources of correlations (Hernández-Fernández, 2011) results into a robust approach to the dependency between the mean size of the parts and the number of parts with lighter prior assumptions.

A relationship between the number of parts and the mean size of the parts consistent with Menzerath-Altmann law has been reported in genomes, not only between chromosome number and mean chromosome size (here; Ferrer-i-Cancho & Forns, 2009), but also between the number of exons of a gene and the mean exon size in the human genome (Li, 2012). As for the second discovery, the hypothesis of trivial power law with a -1 exponent (Solé, 2010) has also been excluded (Li, 2012). Indeed, the finding of Menzerath-Altmann law in genomes is not surprising given the many parallels that have been investigated and established between human language and genomes (Bel-Enguix & Jiménez-Lopez 2011 and references therein). However, the origins and the depth of this statistical coincidence between language, genomes and also music (Boroda & Altmann 1991) should be investigated further.

## 5. METHODS

(a) Data

The same dataset as in the study by Hernández-Fernández et al (2011), which is an updated version of that of Ferrer-i-Cancho & Forns (2009), was used. Group sizes are shown in Table 1.

(b) Non-linear regression.

Throughout the article the fit of the functions defined in Eqs. 2-4 is studied. The goodness of the fit of these equations is evaluated by means of the residual standard error defined as

$$s = \sqrt{\frac{RSS}{N-p}},$$ (5)

where $RSS$ is the residual sum of squares and $N-p$ is the degrees of freedom ($N$ is the sample size and $p$ is the number of parameters; $p=1$ for Eq. 2, $p=2$ for Eq. 3 and $p=3$ for Eq. 4).

The general form of RSS for our functions is

$$RSS = \sum_{i=1}^{N}(y_i - ax_i^b e^{cx_i})^2, \quad (6)$$

where $N$ is the number of organisms of the group and, $x_i$ and $y_i$ are, respectively, the value of $L_g$ and $L_c$ of the $i$-th organism of the group.

The parameters that give the best fit minimizing $s$, which is equivalent to minimizing RSS, are obtained. First, $a^*$, the value of $a$ that minimizes RSS given $b$ and $c$, will be derived. The condition $dRSS/da = 0$ yields

$$a^* = \frac{\sum_{i=1}^{N}(y_i x_i^b e^{cx_i})}{\sum_{i=1}^{N}(x_i^{2b} e^{cx_i})}. \quad (7)$$

$a^*$ corresponds to a minimum of RSS (given $b$ and $c$) if and only if $d^2RSS/d^2a > 0$. In our case, we have

$$\frac{d^2 RSS}{d^2 a} = \sum_{i=1}^{N} x_i^{2b} e^{cx_i} > 0 \quad (8)$$

because the number of chromosomes is a strictly positive number and thus $a^*$ is a minimum.

When $b=1$ and $c=0$, one obtains the well-known estimator of the slope of a linear function through the origin (Sheather, 2010; pp. 41), i.e.

$$a^* = \frac{\sum_{i=1}^{N} y_i x_i}{\sum_{i=1}^{N} x_i^2}, \quad (9)$$

and the condition in Eq. 8 holds trivially regardless of the sign of the $x_i$'s. The best fit of Eq. 2 can be obtained exactly applying $b=-1$ and $c=0$ to Eq. 7, which gives

$$a^* = \frac{\sum_{i=1}^{N}(y_i/x_i)}{\sum_{i=1}^{N}(1/x_i^2)}. \quad (10)$$

For the other functions (Eqs. 3 and 4), a nonlinear regression algorithm (based on the Gauss-Newton method) that minimizes RSS numerically (Ritz & Streibig, 2008) was used. It is well-known that providing the appropriate initial values of the parameters of the ideal function is crucial for the success of the non-linear regression algorithm (Ritz & Streibig, 2008) because the algorithm could get trapped in local minima of RSS. It is customary to apply a transformation of the original curve to express it in a way that simple linear regression can be applied to obtain the initial values of the parameters for the non-linear regression technique (Ritz & Streibig, 2008). For the fit of Eq. 3, the nonlinear regression algorithm was fed with initial values of $a$ and $b$ estimated through a linear regression of Eq. 3 in logarithmic scale (Ritz & Streibig, 2008). Eq. 3 is equivalent to

$$y' = bx' + a', \tag{11}$$

where $y' = \log L_c$, $x' = \log L_g$ and $a' = \log a$. A standard least squares linear regression gives the initial value of $b$ and $a'$. The initial value of $a'$ is obtained through $a = e^{a'}$.

As for the fit of Eq. 4, two different starting values of $a$, $b$ and $c$ were considered and the final values of $a$, $b$ and $c$ that yielded the smallest value of *RSS* where retained. The first initial set up was defined by $a$ and $b$ by the best fit of Eq. 3 obtained through nonlinear regression and $c = 0$. The second initial set up was defined by $b = 0$ and the values of $a$ and $c$ estimated from linear regression on a logarithmic transformation of Eq. 4 with $b = 0$. If logarithms are taken on both sides of Eq. 4, one obtains

$$y' = cL_g + a', \tag{12}$$

where $y' = \log L_c$, and $a' = \log a$ as before. A standard least squares linear regression gives the initial value of $c$ in the second initial setup and $a'$. The initial value of $a'$ in the second initial setup is obtained through $a = e^{a'}$ as before.

The confidence intervals for $b$ shown in Table 1 were computed using a nonparameteric bootstrap approach with 999 artificially generated datasets (Ritz & Streibig, 2008; pp. 96-99). A bootstrap technique was used instead of profile confidence intervals or Wald confidence intervals because of the greater robustness of the bootstrap approach (Ritz & Streibig, 2008).

When evaluating the power of the fit yielded by Eq. 4 versus that of Eq. 3, an extra-sums-of-squares $F$ test was used to determine if parameter $c$ can be neglected and thus Eq. 4 could be reduced to Eq. 4 (Ritz & Streibig, 2008; pp. 103-105). We used an $F$ test instead of a $t$ test because the former is more robust than the latter (Ritz & Streibig, 2008). The same analysis was performed for evaluating the power of the fit of Eq. 3 versus that of Eq. 2.

## ACKNOWLEDGEMENTS

This article is dedicated to G. Altmann and the late P. Menzerath. We thank G. Altmann for his clarifications and J. Perarnau for technical advice on the statistical analyses. This work was supported by the grant *Iniciació i reincorporació a la recerca* from the Universitat Politècnica de Catalunya and the grant BASMATI (TIN2011-27479-C04-03) from the Spanish Ministry of Science and Innovation (RFC and JB).

Table 1. Summary of the fit of $L_c=aL_g^b$. $L_c$ is the mean chromosome length and $L_g$ is the number of chromosomes in the major groups of organisms from the study by Hernández-Fernández et al (2011). The number attached to the group name indicates the number of organisms for that group in our dataset. For the parameters *a* and *b* the notation "estimate ± standard error" is used. $b_{min}$ and $b_{max}$ are, respectively, the lower and the upper bound of *b* in a 97.5% confidence interval. *F* is the value of the *F* statistic used to determine if parameter *b* contributes to decrease error significantly with regard to the error obtained by the fit of $L_c=aL_g^{-1}$. *p* is the p-value of the corresponding *F* test. The values of *F* were rounded to leave only two significant digits. The values of *p* were rounded to leave a single significant decimal. (*) is used to indicate the exponents that are inconsistent with *b* = -1, the prediction of Solé (2010), according to the *F* test at a significance level of 0.05.

| Group | a | b | $b_{min}$ | $b_{max}$ | F | p |
|---|---|---|---|---|---|---|
| Fungi (56) | 11 ± 3 | -0.5* ± 0.2 | -0.8 | -0.2 | 9.9 | 0.003 |
| Angiosperm plants (4706) | (51 ± 5)×10² | -0.95 ± 0.05 | -1.04 | -0.86 | 1.5 | 0.2 |
| Gymnosperm plants (170) | (3 ± 2) ×10³ | -0.3* ± 0.2 | -0.8 | 0.1 | 10 | 0.001 |
| Insects (269) | (6 ± 1)×10² | -0.7* ± 0.1 | -0.9 | -0.5 | 6.0 | 0.01 |
| Reptiles (170) | (8 ± 3)×10² | -0.6* ± 0.1 | -0.8 | -0.4 | 9.2 | 0.003 |
| Birds (99) | (16 ± 5)×10² | -1.0 ± 0.1 | -1.3 | -0.9 | 0.17 | 0.7 |
| Mammals (371) | (33 ± 1)×10² | -0.99 ± 0.02 | -1.02 | -0.96 | 0.82 | 0.4 |
| Cartilaginous fishes (52) | (3 ± 2)×10² | -0.8 ± 0.2 | -1.3 | -0.4 | 0.75 | 0.4 |
| Jawless fishes (13) | (11 ± 2)×10³ | -1.45 ± 0.08 | -1.60 | -1.32 | 41 | 6×10⁻⁵ |
| Ray-finned fishes (647) | (30 ± 9)×10 | -0.54* ± 0.09 | -0.75 | -0.36 | 14 | 2×10⁻⁴ |
| Amphibians (315) | (10 ± 6)×10² | 0.1* ± 0.2 | -0.4 | 0.5 | 21 | 6×10⁻⁶ |

Table 2. Summary of the fit of $L_c = aL_g^b e^{cL_g}$. $L_c$ is the mean chromosome length and $L_g$ is the number of chromosomes in the major taxonomic groups from the study by Hernández-Fernández et al (2011). For the parameters *a, b* and *c* the notation "estimate ± standard error" is used. *F* is the value of the *F* statistic used to determine if parameter *c* contributes to decrease error significantly with regard to the error obtained by the fit of $L_c = aL_g^b$. *p* is the p-value of the corresponding *F* test. The values of *F* were rounded to leave only two significant digits. The values of *p* were rounded to leave a single significant decimal. (*) is used to mark the values of *c* that are significantly different from zero according to the *F* test at a significance level of 0.05.

| Group | a | b | c | F | p |
|---|---|---|---|---|---|
| Fungi | 10 ± 9 | -0.9 ± 0.5 | -0.05 ± 0.06 | 0.76 | 0.4 |
| Angiosperm plants | (28 ± 6)×10$^2$ | -0.36 ± 0.2 | -0.07* ± 0.02 | 20 | 8×10$^{-6}$ |
| Gymnosperm plants | (0.3 ± 7)×10$^{12}$ | 23 ± 5 | -1.9* ± 0.4 | 28 | 2×10$^{-7}$ |
| Insects | (6 ± 3)×10$^2$ | -0.7 ± 0.5 | 0.01 ± 0.07 | 0.019 | 0.9 |
| Reptiles | (0.8 ± 1)×10$^4$ | -1.9 ± 0.7 | 0.07 ± 0.04 | 2.8 | 0.1 |
| Birds | (4 ± 5)×10$^3$ | -1.5 ± 0.6 | 0.03 ± 0.03 | 0.63 | 0.4 |
| Mammals | (30 ± 2)×10$^2$ | -0.89 ± 0.05 | -0.009* ± 0.004 | 4.1 | 0.04 |
| Cartilaginous fishes | (0.4 ± 1)×10$^3$ | 0.1 ± 1 | -0.03 ± 0.04 | 0.78 | 0.4 |
| Jawless fishes | (9 ± 4)×10$^3$ | -1.3 ± 0.3 | -0.009 ± 0.02 | 0.20 | 0.7 |
| Ray-finned fishes | (2 ± 1)×10$^3$ | -1.4 ± 0.2 | 0.023* ± 0.005 | 15 | 10$^{-4}$ |
| Amphibians | (0.6 ± 6)×10$^{16}$ | 26 ± 5 | -1.6* ± 0.4 | 44 | 2×10$^{-10}$ |

Table 3. The error of the fit versus the number of free parameters. A summary of the goodness of the fit of different equations for the relationship between $L_c$, the mean cromosome size and $L_g$, the number of chromosomes. The goodness of the fit is measured with $s$, the residual standard errors. Values were rounded to leave only two decimal digits. (*) is used to indicate the cases where the non-linear regression technique failed when fitting the $n$-parameter equation model in the sense that yielded a value of $s$ higher than that of the $n$-1 parameter equation, with $n$ = 2 or $n$ = 3.

| Group | $s$ | | |
|---|---|---|---|
| | $L_c=aL_g^{-1}$ | $L_c=aL_g^b$ | $L_c=aL_g^b e^{cL_g}$ |
| Fungi | 1.91 | 1.77 | 1.77 |
| Angiosperm plants | 806.69 | 806.64 | 805.02 |
| Gymnosperm plants | 627.37 | 610.39 | 565.20 |
| Insects | 194.62 | 192.83 | 193.18* |
| Reptiles | 38.35 | 37.45 | 37.25 |
| Birds | 9.08 | 9.12 | 9.14* |
| Mammals | 49.76 | 49.77 | 49.56 |
| Cartilaginous fishes | 87.12 | 87.34 | 87.53* |
| Jawless fishes | 55.13 | 26.48 | 27.51* |
| Ray-finned fishes | 27.44 | 27.17 | 26.87 |
| Amphibians | 1418.08 | 1374.87 | 1289.92 |

Table 4. Summary of results on the dependency between $L_c$ and $L_g$. The analysis of the dependency between $L_c$ and $L_g$ in all the groups of organisms through the fit of $L_c = aL_g^b e^{cL_g}$. (*) Is used to indicate results borrowed from (Ferrer-i-Cancho & Forns, 2009) at a significance level of 0.05 (the significant and non-significant correlations are the same if the dataset of the present article is used).

| Group | Correlation between $L_c$ and $L_g$ (*) | $b \neq -1$ when $c=0$ (Table 1) | $c \neq 0$ (Table 2) | $b=-1$ and $c=0$ yield maximum $s$ (Table 3) |
|---|---|---|---|---|
| Fungi | Yes | Yes | | Yes |
| Angiosperm plants | Yes | | Yes | Yes |
| Gymnosperm plants | | Yes | Yes | Yes |
| Insects | Yes | Yes | | Yes |
| Reptiles | Yes | Yes | | Yes |
| Birds | Yes | | | |
| Mammals | Yes | | Yes | |
| Cartilaginous fishes | Yes | | | |
| Jawless fishes | Yes | Yes | | Yes |
| Ray-finned fishes | | Yes | Yes | Yes |
| Amphibians | Yes | Yes | Yes | Yes |